\newif\ifShowKeys
\ShowKeystrue
\ShowKeysfalse

\documentclass[11pt]{article}
	\pdfoutput=1
\usepackage[usenames,dvipsnames]{xcolor}
\usepackage[setpagesize=false,pagebackref=false, linktocpage, bookmarksopen=true, colorlinks=true, linkcolor=Maroon,citecolor=Maroon,urlcolor=Maroon]{hyperref}
\usepackage[parsep]{collref}

\ifShowKeys \usepackage[notcite]{showkeys} \fi

\usepackage{amsmath, amssymb,amsthm}
\numberwithin{equation}{section}

\usepackage{bm,environ,mathrsfs,array,arydshln}
\usepackage{booktabs,float,slashed,hyperref}
\usepackage[mathcal]{euscript}
\usepackage{tensor} 						% Ratcliffe package to write tensors
\usepackage{mathabx}
\usepackage[vcentermath]{youngtab}
\usepackage{xcolor}

\usepackage{aurical}
\usepackage[T1]{fontenc}
\usepackage[nodayofweek]{date time}
\usepackage{graphicx,epsfig,epic}
\usepackage{tikz} 
\usetikzlibrary{arrows,decorations.pathreplacing,decorations.markings,snakes}
\usetikzlibrary{cd}

\usepackage{framed}						% for shaded equations \begin{shaded}...\end{shaded}
\definecolor{shadecolor}{rgb}{0.9996078, 0.984314, 0.960784}

\usepackage{comment}

\allowdisplaybreaks

\definecolor{myred}{RGB}{233, 33, 45}

\newcommand{\bs}{\begin{shaded}}
\newcommand{\es}{\end{shaded}\noindent}
\def\ba#1\ea{\begin{align}#1\end{align}}		% very clever way to bypass the known problem...
\newcommand{\be}{\begin{equation}}
\newcommand{\ee}{\end{equation}}
\newcommand{\bea}{\begin{equation} \begin{aligned}} 
\newcommand{\eea}{\end{aligned} \end{equation}}
\newcommand{\mc}{\mathcal }

\newcommand{\la}{\label}
\newcommand{\eps}{\varepsilon}

\newcommand{\lp}{\notag \\ & }

\newcommand{\wt}{\widetilde}

\newcommand{\cf}{\textit{cf.} }
\newcommand{\ie}{\textit{i.e.} }

\newcommand{\N}{\mathcal N}

\renewcommand{\l}{\lambda}

\DeclareMathOperator{\Tr}{Tr}
\DeclareMathOperator{\PE}{PE}

\newcommand{\I}{\mathrm{I}}

\newcommand{\vth}{\vartheta}
\newcommand{\z}{\rho}
\newcommand{\bz}{\bm{z}}
\newcommand{\bg}{\bm{g}}
\newcommand{\bQ}{\bm{Q}}
\newcommand{\wh}{\widehat}

\newcommand{\smb}{\scalebox{0.6}{$\Box$}}

\begin{document}

%\date{\currenttime}
%\begin{flushright}\boxed{\small{\tt \today \ \ - \ \  \currenttime }}\end{flushright}

%\centerline{\Large\sc  Schur index and D3 branes - Notes}
%\vskip 0.5cm
%% \centerline{\sc M. Beccaria}
%\bigskip
%\begin{abstract}
%\begin{center}
%%\includegraphics[width=0.5\textwidth]{cover}
%\end{center}
%\end{abstract}

%%%%%%%%%%%%%%%%%%%%%%%%%%%%%%%%%%%%%%%%%%%%%%%%%%%%%%%%%
%%%%% NOTICE the title page is commented out with \begin{comment}...\end{comment}   but is ready to be used
%\begin{comment}

\begin{titlepage}
%\begin{tabbing}
%\hspace*{11.5cm} \=  \kill % set the tabbings
%\>  Imperial-TP-AT-2024-?? \\
%\> %none
%\end{tabbing}

\vspace*{15mm}
\begin{center}
{\Large\sc   Schur line defect correlators}\vskip 9pt
{\Large\sc      and giant graviton expansion}

\vspace*{10mm}

{\Large M. Beccaria}
%\footnote{\ Also at the Institute for Theoretical and Mathematical Physics (ITMP) of Moscow University   and Lebedev Institute.}} 

\vspace*{4mm}
	
${}^a$ Universit\`a del Salento, Dipartimento di Matematica e Fisica \textit{Ennio De Giorgi},\\ 
		and I.N.F.N. - sezione di Lecce, Via Arnesano, I-73100 Lecce, Italy
			\vskip 0.3cm
%${}^b$ Blackett Laboratory, Imperial College London SW7 2AZ, U.K.
%			\vskip 0.3cm
\vskip 0.2cm {\small E-mail: \texttt{matteo.beccaria@le.infn.it}}
\vspace*{0.8cm}
\end{center}

\begin{abstract}  
We consider  Schur line defect correlators in four dimensional $\N=4$ $U(N)$ SYM and  their giant graviton expansion encoding 
finite $N$ corrections to the large $N$ limit.
We compute in closed form the single giant graviton contribution to  correlators with general insertions of 
$\frac{1}{2}$-BPS charged Wilson lines. For the 2-point function with fundamental and anti-fundamental Wilson lines, we match the 
result from fluctuations of two half-infinite strings ending on the giant graviton, recently proposed in arXiv:2403.11543.
In particular, we prove  exact factorization of the defect contribution with respect to wrapped D3 brane fluctuations 
representing  the single giant graviton correction to the undecorated
Schur index. This follows from a finite-difference representation of the Schur line defect index in terms of the index without defects,
and similar factorization holds quite generally for more complicated defect configurations. In particular, 
the single giant graviton contribution to the 4-point function with two fundamental and two anti-fundamental lines is computed and discussed in this 
perspective.
\end{abstract}
\vskip 0.5cm
	{
		%Keywords: {\sc insert here keywords}
	}
\end{titlepage}
%\end{comment}
%%%%%%%%%%%%%%%%%%%%%%%%%%%%%%%%%%%%%%%%%%%%%%%%%%%%%%%%%

\tableofcontents
\vspace{1cm}

%\section*{To be done}

%\section*{\red{Questions}}
%
%\begin{enumerate}
%\item
%\end{enumerate}

\section{Introduction}

The superconformal index introduced in \cite{Kinney:2005ej,Romelsberger:2005eg,Romelsberger:2007ec} is the 
Witten index \cite{Witten:1982df}  in radial quantization and is a common  device for the study of the BPS
spectrum of superconformal theories.
As a general fact, for $U(N)$ gauge  theories with a gravity dual, the superconformal index has a definite $N\to \infty$ limit 
matching the index of BPS
supergravity states. \footnote{This is a limit order by order in the charge of the contributing BPS states. It captures single trace states in the CFT, 
but should not be confused with the large $N$ limit of the index, see in particular \cite{Agarwal:2020zwm,Murthy:2020rbd}.
}
On the CFT side, corrections at finite $N$ are due to gauge group trace relations taking into account the structure of multi-trace states. 
The dual gravity explanation of these corrections is in terms of 
giant graviton contributions \cite{McGreevy:2000cw} with charge $\sim N$. 
As suggested recently in  \cite{Chang:2024zqi, DeddoLiu}, finite $N$ corrections may also be computed by analyzing the 
BPS geometries of supergravity bubbling solutions.

In the specific case of 4d $\N=4$ $U(N)$ SYM, dual to IIB superstring in $AdS_{5}\times S^{5}$, the giant graviton expansion of the index
involves  the ``brane index'' of the theory living on the world-volume of multiply wrapped D3 brane configurations \cite{Imamura:2021ytr,Gaiotto:2021xce,Lee:2022vig}.
In this paper, we will consider the Schur specialization of the general index \cite{Gadde:2011ik,Gadde:2011uv}. It may be introduced in 
theories with at least $\N=2$ supersymmetry where it reproduces  the vacuum character of the chiral algebra characterizing a protected sector \cite{Beem:2013sza}. 
In $\N=4$ $U(N)$ SYM the definition of the Schur index is 
\be
\I^{U(N)}(\eta; q) = \Tr_{\rm BPS}[(-1)^{\rm F}\, q^{H+J+\bar J}\eta^{R_{1}-R_{2}}]\, ,
\ee
where  $H$ is the Hamiltonian, $J, \bar J$ are two spins, and the $R_{1}, R_{2}$ are two of the R-charge generators in the $PSU(2,2|4)$ superconformal group.
The variable $q$ is the universal fugacity, while $\eta$ is usually referred to as a flavor fugacity. 
The Schur index admits an  explicit holonomy integral representation that reads (PE stands for plethystic exponentiation), see for instance \cite{Gaiotto:2019jvo}, 
\ba
\la{14}
\I^{U(N)}(\eta; q) & = {\oint}_{|\bz|=1}D^{N}\bz\, \PE[ f(\eta; q)\chi_{\smb}(\bz)\chi_{\smb}(\bz^{-1})]\,,
\ea
where the measure $D^{N}\bz$ and the character $\chi_{\smb}(\bz)$ of the $U(N)$ fundamental representation are 
\be
D^{N}\bz = \frac{1}{N!}\prod_{n=1}^{N}\frac{dz_{n}}{2\pi i\, z_{n}}\prod_{n\neq m}\bigg(1-\frac{z_{n}}{z_{m}}\bigg), \qquad
\chi_{\smb}(\bz) = \sum_{n=1}^{N}z_{n}\,.
\ee
The function $f(\eta; q)$ in (\ref{14}) is the single particle Schur index and is given by the simple function
\be
f(\eta; q) = \frac{(\eta+\eta^{-1})\, q-2q^{2}}{1-q^{2}}\,.
\ee
Exact results for the Schur index at specific values of $N$
have been obtained in \cite{Bourdier:2015sga,Bourdier:2015wda,Pan:2021mrw,Hatsuda:2022xdv} in the case of $U(N)$ gauge group, 
and generalized to $B_{n}$,$C_{n}$, $D_{n}$, $G_{2}$  groups in  \cite{Du:2023kfu}.

As discussed in \cite{Arai:2020qaj}, the giant graviton expansion of the Schur index is simpler than that of the general index and 
can be  written in terms of the $\N=4$ SYM index itself. One has indeed 
the remarkable relation 
\ba
\la{1.2}
\I^{U(N)}(\eta; q) = \I^{\rm KK}(\eta;q)\,  \sum_{n=0}^{\infty}\sum_{p=0}^{n} (\eta q)^{(n-p)N}\I^{\rm D3}_{n-p}(\eta; q) q^{2(n-p)p}(\eta^{-1}q)^{pN}
\I^{\rm D3}_{p}(\eta^{-1}; q)\, ,
\ea
where $\I^{\rm KK}(\eta;q)$ is the large $N$ Kaluza-Klein supergravity contribution and the brane indices $\I^{\rm D3}_{n}(\eta; q)$ are obtained
by analytic continuation of the $\N=4$ $U(n)$ SYM index \cite{Arai:2019xmp}
\be
\la{1.3}
\I^{\rm D3}_{n}(\eta; q) = \I^{U(n)}(\eta^{-1/2}q^{-3/2}; \eta^{-1/2}\,q^{1/2})\,.
\ee
The terms in the sum (\ref{1.2}) are organized in contributions with weight $\sim q^{nN}$ with the index $n$ being the wrapping number of  
D3 branes with topology $S^{1}\times S^{3}$, where $S^{1}\subset AdS_{5}$ and $S^{3}\subset S^{5}$. 
The approach based on the analytic continuation (\ref{1.3}) was successfully applied
in many other instances \cite{Arai:2019wgv,Arai:2019aou,Arai:2020uwd,Fujiwara:2021xgu,Imamura:2021dya,Fujiwara:2023bdc,Imamura:2022aua,Fujiwara:2023bdc}
and was confirmed by complete fluctuation analysis in \cite{Beccaria:2023sph,Beccaria:2023cuo,Beccaria:2024vfx,Gautason:2024nru}. 
\footnote{An alternative approach 
based on localization of the theory on the brane world-volume has been recently discussed in \cite{Lee:2023iil} and 
later developed in \cite{Eleftheriou:2023jxr} for the giant graviton expansion of the 
$\frac{1}{2}$-BPS  index in $\N=4$ $U(N)$ SYM. 
}
Recently, the brane indices $\I_{n}^{\rm D3}(\eta; q)$ were computed in closed form in \cite{Beccaria:2024szi}.

A natural extension of the Schur index consists in decorating it by inserting defect lines \cite{Dimofte:2011py,Gang:2012yr}
and  exact results have been obtained 
for the associated Schur correlators
involving an arbitrary number of defect operator ('t Hooft or Wilson lines) insertions \cite{Drukker:2015spa,Cordova:2016uwk,Neitzke:2017cxz,Hatsuda:2023iwi,Hatsuda:2023imp,Hatsuda:2023iof,Guo:2023mkn}.  
When the index is regarded as a supersymmetric partition function on $S^{1}\times S^{3}$, the defect lines are wrapping $S^{1}$ and 
are placed on a great circle of $S^{3}$ to preserve supersymmetry \cite{Cordova:2016uwk}. Schur line defect correlators are topological 
and do not depend on the distance between the inserted Wilson lines. Here, we consider the insertion of $\frac{1}{2}$-BPS Wilson lines with generic charges.

Insertion of Wilson line defects in representations $R_{1}, R_{2}, \dots$ is computed by the following modification of the holonomy integral in (\ref{14})
\ba
\la{1.7}
\I^{U(N)}_{R_{1}, R_{2}, \dots}(\eta; q) & = {\oint}_{|\bz|=1}D^{N}\bz\, \prod_{n\ge 1}\chi_{R_{n}}(\bz)\, 
\PE[ f(\eta; q)\chi_{\smb}(z)\chi_{\smb}(z^{-1})]\,.
\ea
Due to its basic role in the following discussion, we introduce a  special notation for the Schur line defect 2-point function with a fundamental and an anti-fundamental
\be
\I^{U(N)}_{\rm F}(\eta; q)\equiv \I^{U(N)}_{\smb, \overline{\smb}}(\eta; q) = \I^{U(N)}_{\rm adjoint}(\eta; q)\,.
\ee
When the Wilson lines  in fundamental representation have charges $Q_{n}$, the corresponding expression involves multiple power symmetric characters 
$\prod_{n}\chi_{\smb}(\bz^{Q_{n}})$.

In this paper, we consider the giant graviton expansion of the above Schur line defect correlators, working at  single giant graviton level. 
In the case of the 2-point function $\I_{\rm F}^{U(N)}(\eta; q)$, the  large $N$ limit is known to take the factorized form \cite{Gang:2012yr}
\be
\la{x19}
\I_{\rm F}^{U(\infty)}(\eta; q) = \I_{\rm F1}(\eta; q)\times \I^{U(\infty)}(\eta; q), \qquad  \I_{\rm F1}(\eta; q) = \frac{1}{1-f(\eta; q)}\,.
\ee
From the AdS/CFT perspective, the factor $\I_{\rm F1}(\eta; q)$   corresponds to fluctuations of a fundamental string along $AdS_{2}\subset AdS_{5}$ \cite{Rey:1998ik,Maldacena:1998im}
 meeting the boundary of $AdS_{2}$ at the two poles of $S^{3}$
in $\partial AdS_{5} = \mathbb{R}\times S^{3}$, where the line operators are placed. The detailed analysis of fluctuations was performed  in 
\cite{Gang:2012yr}  \footnote{Fluctuations are in multiplets of $SO(2,1)\times SO(3)\times SO(5)$. The last two factors 
correspond to rotations of the remaining $AdS_{5}$ coordinates giving $SO(3)$ symmetry, and rotations of the fixed point in $S^{5}$ giving $SO(5)$ \cite{Faraggi:2011bb}.}
confirming that the expression $f_{\rm F1}(\eta; q) = -q^{2}+(\eta+\eta^{-1})\, q$ in the formula
\be
\I_{\rm F1}(\eta; q) = \PE[f_{\rm F1}(\eta; q)]\, , 
\ee
matches the single particle index of fluctuations of the fundamental string.

Quite naturally, one expects that for the 2-point function $\I_{\rm F}^{U(N)}(\eta; q)$ one should also have a leading giant graviton contribution 
due to the contribution from two semi-infinite strings attached to the Wilson lines and ending on the giant graviton. This proposal and its quantitative verification
 appeared very recently
 in  \cite{Imamura:2024lkw} (with extension to  multi-graviton contributions). In particular, it was found that \footnote{In the ratio (\ref{ratio}) we subtract 
 the contribution $\I_{\rm F1}\I^{(N)}$
 corresponding to the case when the two half-infinite strings do not end on the giant graviton, and the difference is divided by the supergravity contribution.
 }
\ba
\la{ratio}
& \frac{\I_{\rm F}^{U(N)}-I_{\rm F1}\,\I^{U(N)}}{\I^{U(\infty)}}  = 
1+\bigg(\mc G^{+}_{\rm F}(\eta; q)\, \eta^{N}+\mc G^{-}_{\rm F}(\eta; q)\,\eta^{-N}\bigg)\, q^{N}+\mc O(q^{2N})\,, \\
\la{1.12}
& \mc G^{+}_{\rm F}(\eta; q)  =  G^{+}_{\rm D3}(\eta; q)\times \frac{1}{\eta q}\,  \PE[f_{\rm F}(\eta; q) ]\, , 
\qquad f_{\rm F}(\eta; q) = 2\eta^{-1}q-2q^{2}\, ,
\ea
where  $G^{+}_{\rm D3}(\eta; q)$ is the single giant graviton contribution 
to the undecorated Schur index coming from D3 brane fluctuations \cite{Arai:2020qaj,Beccaria:2024szi} and $f_{\rm F}(\eta; q)$ 
 in (\ref{1.12}) agrees with the single particle index from fluctuations of the two semi-infinite strings 
ending on the giant graviton. 
The origin of the prefactor $1/(\eta q)$ is at the moment unclear. It also affects higher order giant graviton contributions and was suggested to be a back-reaction effect in 
\cite{Imamura:2024lkw}.  Relations (\ref{ratio}, \ref{1.12}) were confirmed  by comparing with the first terms in the small $q$ expansion of the 2-point function
at large $N$.

In this paper, we derive the exact form of the single giant graviton expansion of various Schur line defect correlators. In particular, for 
$\I^{U(N)}_{\rm F}(\eta; q)$ we  prove the exact result
\ba
\la{x113}
\frac{\I^{U(N)}_{\rm F}(\eta; q)}{\I^{U(\infty)}_{\rm F}(\eta; q)} = 1+\bigg(G^{+}_{\rm F}(\eta; q)\, \eta^{N}+G^{-}_{\rm F}(\eta; q)\,\eta^{-N}\bigg)\, q^{N}+\mc O(q^{2N}) \,,
\ea
with (see Appendix \ref{app:special} for notation)
\be
\la{1.10}
G^{+}_{\rm F}(\eta; q) = -\eta^{2}\,q\,\bigg(
1+\frac{1-q^{2}}{\eta\, q}\frac{1-\eta\, q}{1-\eta^{-1}\, q}
\bigg)\frac{(\frac{q}{\eta})_{\infty}^{3}}{\vth(\eta^{2},\frac{q}{\eta})}, \qquad G^{-}_{\rm F}(\eta; q) = G^{+}_{\rm F}(\eta^{-1}; q).
\ee

Expression (\ref{1.10}) is equivalent to (\ref{ratio}, \ref{1.12}) and has a remarkable factorized form. Indeed, the effect of the two Wilson lines insertion
is fully captured by the second term in round bracket. In other words, one has the exact simple relation 
\be
\la{115}
G^{+}_{\rm F}(\eta; q) = \bigg(
1+\frac{1}{\eta\, q}\frac{(1-q^{2})(1-\eta\, q)}{1-\eta^{-1}\, q}
\bigg)\, G^{+}_{\rm D3}(\eta; q)\, .
\ee
The correction factor in brackets has the factor $1/(\eta q)$,  as in (\ref{1.12}), while the rest coincides with the  two
half  strings fluctuations.
Our derivation  builds on the results of \cite{Murthy:2022ien} for general multi-coupling unitary matrix models.
We will illustrate how to reduce the calculation of the Schur line defect correlators to finite differences of the undecorated Schur index with respect to the gauge group rank.
The factorization property in (\ref{115}) will then follow as a simple consequence.

By our approach, it will be possible to generalize results like (\ref{115}) to a large extent. To give an example, for the 4-point function with 
two Wilson lines in the fundamental representation and two in the anti-fundamental, we obtain for the ratio similar to (\ref{x113}) the exact result 
\ba
G_{\smb, \smb, \overline{\smb}, \overline{\smb}}^{+}(\eta; q) =  \bigg[
1+\frac{1-5q^{2}+3\eta q+\eta q^{3}}{2\eta^{2}\, q^{2}}\frac{(1-q^{2})(1-\eta\, q)}{(1-\eta^{-1}\, q)^{2}}
\bigg]\, G_{\rm D3}^{+}(\eta; q)\, .
\ea
For the subtracted ratio similar to (\ref{ratio}), this gives 
\ba
\mc G^{+}_{\smb, \smb, \overline{\smb}, \overline{\smb}}(\eta; q)  = G_{\rm D3}^{+}(\eta; q)\times 
\frac{1-5q^{2}+3\eta q+\eta q^{3}}{\eta^{2} q^{2}}\, 
\PE[-3q^{2}+4\eta^{-1}q+\eta q ]\, .
\ea
Comparing with (\ref{1.12}), we see that it takes a factorized form with a more complicated prefactor which is however still a sum of monomials, times
 plethystic of a three terms combination. This should be the single particle index for fluctuations of the worldsheet attached to the four Wilson lines and ending on the giant graviton.

Our methods may provide  exact predictions for many Schur line defect correlators to be hopefully compared with the analysis of explicit string fluctuations.

\paragraph{Plan of the paper} 

The plan of the paper is the following. In Section \ref{sec:part}, we discuss the large $N$ limit of Schur line defect correlators.
 In Section \ref{sec:main}, we present our main results. In particular, we derive the exact result (\ref{1.10})
for the Schur defect  2-point function with two Wilson lines in the fundamental and anti-fundamental. The derivation includes the case of 
a pair of oppositely charged Wilson lines.
In  Section \ref{sec:four} we obtain the single giant graviton correction to the  4-point function with two fundamental
and two anti-fundamental lines.  Section \ref{sec:gravity} discusses the relation between our closed formulas and their interpretation in terms
of string fluctuations. The case of general charge assignments with a vanishing large $N$ limit, but admitting a non-trivial
single giant graviton correction, is presented in  Appendix \ref{app:zero}.

\section{Schur line defect correlators and the large $N$ limit}
\la{sec:part}

In this section, we begin by discussing  the $N\to \infty$ limit of Schur line defect correlators.
Let us start from the multi-coupling unitary matrix integral \cite{Murthy:2022ien} with $\bg = (g_{1}, g_{2}, \dots)$
\be
\la{2.1}
Z_{N}(\bg) = \int_{U(N)}dU\, \exp\bigg(\sum_{n=1}^{\infty}\frac{1}{n}g_{n}\Tr U^{n}\Tr U^{-n}\bigg)\,.
\ee
The Schur index is obtained by  specialization
\be
\la{22}
\I^{U(N)}(\eta; q) = Z_{N}(\bg), \qquad g_{n} = f(\eta^{n}; q^{n})\,.
\ee
An important result of  \cite{Murthy:2022ien} is the following large $N$ limit
\be
Z_{\infty}(\bg) =  \prod_{n=1}^{\infty}\frac{1}{1-g_{n}}\,.
\ee
From this result, we can obtain the large $N$ limit of correlators with any number of pairs of oppositely charged Wilson lines. For instance, we have 
\be
\la{x24}
Z_{N}^{\rm F}(\bg) = \frac{\partial}{\partial g_{1}}Z_{N}(\bg)\,, \qquad\to \qquad
Z_{\infty}^{\rm F}(\bm{g}) =  \frac{1}{1-g_{1}}\prod_{n=1}^{\infty}\frac{1}{1-g_{n}}\,,
\ee
which agrees with (\ref{x19}). Its  string derivation  from fluctuations of a fundamental string along $AdS_{2}\subset AdS_{5}$ was given in 
\cite{Gang:2012yr}. 

Similar expressions can be obtained in more general cases by repeated differentiation. 
The fact that other charge assignments have vanishing  large $N$ limit can be proved by 
group representation theory. An alternative direct derivation is possible by the methods of \cite{Murthy:2022ien}. To this aim, 
let us consider insertions of multiple Wilson lines with arbitrary charges and the Schur line defect correlator
\ba
\I^{U(N)}_{\bQ}(\eta; q) & = \oint_{|\bz|=1}D^{N}\bz\  
\prod_{n=1}^{\infty}\chi_{\smb}(\bz^{q_{n}})\
\PE[ f(\eta; q)\chi_{\smb}(z)\chi_{\smb}(z^{-1})]\,, \\
\bQ &= (q_{1}, q_{2}, \dots), \qquad \sum_{n}q_{n}=0\, .
\ea
Following  \cite{Murthy:2022ien}, we introduce the generating functional
\be
\la{x27}
\wt Z_{N}(\bm{t}^{+}, \bm{t}^{-}) = \int_{U(N)}dU\, \exp\bigg(\sum_{n=1}^{\infty}\frac{1}{n}(t_{n}^{+}\Tr U^{n}+t_{n}^{-} \Tr U^{-n})\bigg)\,.
\ee
For a function $f(\bm{t}^{+}, \bm{t}^{-})$, we define
\be
\la{x28}
\langle f\rangle_{\bm{g}} = \prod_{n=1}^{\infty}\int\frac{dt^{+}_{n}dt^{-}_{n}}{2\pi\, n g_{n}}e^{-\frac{1}{ng_{n}}t^{+}_{n}t^{-}_{n}}\, f(\bm{t}^{+}, \bm{t}^{-})\,,
\qquad 
\int\frac{dt^{+}dt^{-}}{2\pi g}e^{-\frac{1}{g}t^{+}t^{-}}(t^{+})^{n_{+}}(t^{-})^{n_{-}} = n!\, g^{n}\, \delta_{n^{+}, n^{-}}\,.
\ee
The relation between $Z_{N}(\bg)$ and 
$\wt Z_{N}(\bm{t}^{+}, \bm{t}^{-})$
is 
\be
Z_{N}(\bm{g}) = \langle \wt Z_{N}(\bm{t}^{+}, \bm{t}^{-})\rangle_{\bm{g}}\,.
\ee
The $N\to\infty$ limit of the generating functional $\wt Z_{N}$ is  
\be
\wt Z_{\infty}(\bm{t}^{+}, \bm{t}^{-}) = \exp\bigg(\sum_{n=1}^{\infty}\frac{1}{n}t_{n}^{+}t_{n}^{-}\bigg)\,.
\ee
For a set of charges $\bQ = (q_{1}, q_{2}, \dots; -q'_{1}, -q'_{2}, \dots)$, $q_{i},q'_{j}>0$,  we have 
\be
\la{x211}
\wt Z_{\infty}^{\bQ}(\bm{t}^{+}, \bm{t}^{-}) =D_{\bQ} \exp\bigg(\sum_{n=1}^{\infty}\frac{1}{n}t_{n}^{+}t_{n}^{-}\bigg)\, , \qquad
D_{\bQ} = \prod_{i,j}q_{i}q'_{j}\partial_{t^{+}_{q_{i}}}\partial_{t^{-}_{q'_{j}}}\, .
\ee
Thus, 
\ba
Z_{\infty}^{\bQ}(\bg) &=\langle D_{\bQ}\wt Z_{\infty}^{\bQ}(\bm{t}^{+}, \bm{t}^{-}) \rangle_{\bg} 
= \int \prod_{k=1}\frac{dt_{k}^{+} dt_{k}^{-}}{2\pi k g_{k}}\, \exp\bigg(-\sum_{k=1}^{\infty}\frac{1}{kg_{k}}t_k^{+}t_{k}^{-}\bigg)\ D_{\bQ} 
\exp\bigg(\sum_{k=1}^{\infty}\frac{1}{k}t_{k}^{+}t_{k}^{-}\bigg)\, .
\ea
If $\bQ$ is not made of pairs of opposite charges, \ie is not symmetric under a change of sign of all charges, 
we cannot end up with contributions with the same number of $t^{+}_{k}$ and $t_{-}^{k}$ and we get zero due to 
(\ref{x28}). This, together with the previous discussion of the opposite charge cases, proves  the 
conjectures in Section 5.1.1 of \cite{Hatsuda:2023iwi}.

We remark that for general charges $\bQ$ not of the form $(q_{1}, q_{2}, \dots; -q_{1}, -q_{2}, \dots)$, the fact that $Z_{\infty}^{\bQ}(\bg)=0$ means that 
the expression of  $Z_{N}^{\bQ}(\bg)$  starts with a term which is $q^{N}$ times a non-trivial function of $\eta$ and $q$ that may be computed as discussed in 
Appendix \ref{app:zero}.

\section{Leading giant graviton correction to the $(\Box, \overline{\Box})$ 2-point function}
\la{sec:main}

Let us now move to the finite $N$ corrections to the 2-point function in the fundamental, \ie $\I_{\rm F}^{U(N)}(\eta; q)$.
We begin by recalling what is known in the case of the undecorated Schur index. 
Its  leading giant graviton expansion  was derived in closed form in \cite{Beccaria:2024szi}. It reads
(see Appendix \ref{app:special} for our conventions) \footnote{\la{foot1}
Here and in the following, we will denote by $\mc O(q^{2N})$ the double giant graviton contribution. Strictly speaking, its $q$ expansion starts at $q^{2N+\delta}$
for some integer $\delta$. We will not discuss it and just split out all contributions with an explicit $q^{2N}$ factor.}

\be
\la{3.1}
\frac{\I^{U(N)}(\eta; q)}{\I^{U(\infty)}(\eta; q)} = 
1 -\bigg[\eta^{N+2}\frac{(\frac{q}{\eta})_{\infty}^{3}}{\vth(\eta^{2},\frac{q}{\eta})}
+\eta^{-N-2}\frac{(\eta\,q)_{\infty}^{3}}{\vth(\eta^{-2},\eta\,q)}\bigg]\,q^{N+1}+\mc O(q^{2N})\,.
\ee
The first terms of its expansion in small $q$  are
\be
\la{3.2}
\frac{\I^{U(N)}(\eta; q)}{\I^{U(\infty)}(\eta; q)} = 
1 +\bigg[
\frac{\eta}{1-\eta^{2}}(\eta^{N+2}-\eta^{-N-2})+\frac{1}{\eta}(1-\eta^{2})(\eta^{N+1}-\eta^{-N-1})\,q+\cdots\bigg]\, q^{N+1}+\mc O(q^{2N})\,.
\ee

\subsection{Finite $N$ analysis of the holonomy matrix integrals}

Let us examine the structure of the $N$ dependence of $\I_{\rm F}^{U(N)}(\eta; q)$ by  computing explicitly the associated matrix integrals at finite $N$. 
We computed explicit series expansions
at order $q^{N+1}$ for various $N$ and results are collected in Appendix \ref{app:series}. From the pattern guessed in (\ref{B.12}), we have 
\ba
\la{3.3}
\frac{\I^{U(N)}_{\rm F}(\eta; q)}{\I^{U(\infty)}_{\rm F}(\eta; q)} = 1+\bigg[\frac{\eta}{1-\eta^{2}}(\eta^{N+1}-\eta^{-N-1})
+\frac{1-\eta^{2}+\eta^{4}}{\eta(1-\eta^{2})}(\eta^{N}-\eta^{-N}) \,q+\cdots\bigg]\,q^{N}+\mc O(q^{2N})\,.
\ea
This can be written 
\ba
\la{3.4}
\frac{\I^{U(N)}_{\rm F}(\eta; q)}{\I^{U(\infty)}_{\rm F}(\eta; q)} = 1+\bigg(G^{+}_{\rm F}(\eta; q)\, \eta^{N}+G^{-}_{\rm F}(\eta; q)\,\eta^{-N}\bigg)\, q^{N}+\mc O(q^{2N}) \,,
\ea
with 
\bea
\la{3.5}
G_{\rm F}^{+}(\eta; q) &=\frac{\eta^{2}}{1-\eta^{2}}+\frac{1-\eta^{2}+\eta^{4}}{\eta(1-\eta^{2})}\, q+\frac{1-\eta^{4}-2\eta^{6}+\eta^{8}}{\eta^{4}(1-\eta^{2})}\,q^{2}+\mc O(q^{3}), \\
G_{\rm F}^{-}(\eta; q) &= G^{+}_{\rm F}(\eta^{-1}; q)\,. 
\eea
Comparing (\ref{3.3}) with (\ref{3.2}), we notice that the insertion of the adjoint character in the Schur index has the effect of giving a first 
giant graviton correction that starts at order $\mc O(q^{N})$ instead of $\mc O(q^{N+1})$.  

In the next section, we will compute the first term in (\ref{3.5}) which turns out to a simple calculation. This will confirm the pattern conjectured in (\ref{B.12})
and leading to (\ref{3.4}), (\ref{3.5}).
A full calculation of the function $G_{\rm F}^{+}(\eta; q)$ will be presented later.

As a remark, in the unflavored limit $\eta\to 1 $ limit,  we get from (\ref{3.2})
\be
\la{3.6}
\frac{\I^{U(N)}(1; q)}{\I^{U(\infty)}(1; q)} = 1-(N+2)\,q^{N+1}+\mc O(q^{2N})\,,
\ee
in agreement with the exact results in \cite{Bourdier:2015wda}. For the Schur line defect, the unflavored limit can also be read  from the above expressions
and takes the form 
\ba
\la{3.7}
\frac{\I^{U(N)}_{\rm F}(1; q)}{\I^{U(\infty)}_{\rm F}(1; q)} = 1-\bigg[N+1+N\,q-(N+3)\,q^{2}+\cdots\bigg]\, q^{N}+\mc O(q^{2N})\,.
\ea
We will see that this result is actually exact, \ie there are no higher order corrections in the square bracket beyond the shown three terms. The explicit factors of $N$ in (\ref{3.7})
are somehow expected as a general feature of unrefined indices with algebraic constraints on fugacities.  This was discussed as a wall-crossing effect
in \cite{Gaiotto:2021xce,Lee:2022vig,Beccaria:2023zjw}. On gravity side, these factors come from zero modes of wrapped branes fluctuations
 \cite{Beccaria:2023cuo,Beccaria:2024vfx,Gautason:2023igo}.

\subsection{Determination of the $q^{N}$ correction from Young tableaux expansion}

The result 
\ba
\la{x38}
\frac{\I^{U(N)}_{\rm F}(\eta; q)}{\I^{U(\infty)}_{\rm F}(\eta; q)} = 1+\bigg[\frac{\eta}{1-\eta^{2}}(\eta^{N+1}-\eta^{-N-1})+\mc O(q)\bigg]\,q^{N}+\mc O(q^{2N})\,,
\ea
may be obtained in a straightforward way by  Young tableaux expansion methods \cite{Dolan:2007rq}. We illustrate this approach because of its simplicity and
general applicability. It may compute systematically the corrections in (\ref{x38}), 
but we will not delve into this extension, since we will later resum the full set of contributions by a different method.

Introducing  holonomies $\bz = (z_{1},\dots z_{n})$, we have 
\ba
Z_{N}(\bg) &=  \oint_{|\bz|=1}D^{N}\bz\ 
\exp\bigg(\sum_{n=1}^{\infty}\frac{1}{n}g_{n}\chi_{\smb}(\bz^{n})\chi_{\smb}(\bz^{-n})\bigg)\,.
\ea
For two symmetric functions $a(\bm{z}), b(\bm{z})$, following \cite{Dolan:2007rq},  we define the product
\be
\langle a,b\rangle_{N} =\oint_{|\bz|=1}D^{N}\bz\  a(\bm{z})\, b(\bm{z}^{-1})\,.
\ee
A partition $\l$ can be represented as $(\l_{1}, \l_{2}, \dots)$ with $\l_{1}\ge \l_{2} \ge \cdots$
or in frequency representation $1^{r_{1}}2^{r_{2}}\cdots$. The number of parts of $\l$ is $\ell(\l) =\sum_{n}r_{n}$ (the number of non-zero $\l_{i}$). It is the number of rows
in the associated Young tableau. The weight of the partition $\l$ is $|\l| = \sum_{n}\l_{n} = \sum_{n}n r_{n}$, the number of blocks in the Young tableau. 
The known relation between plethystic and Young tableaux gives the expansion 
\be
\la{2.5}
Z_{N}(\bg) = \sum_{d=0}^{\infty}\sum_{|\l|=d}\frac{\bg^{\l}}{\z_{\l}}\ \langle \chi_{\smb}^{\l}(\bz), \chi_{\smb}^{\l}(\bz)\rangle_{N}\,,
\ee
where
\be
\la{2.6}
\bg^{\l} = \prod_{n=1}^{\infty}g_{\l_{n}}= \prod_{n=1}^{\infty}g_{n}^{r_{n}}, \qquad \z_{\l} = \prod_{n=1}^{\infty}r_{n}!\, n^{r_{n}}, \qquad 
\phi^{\l}(\bz) = \prod_{n=1}^{\infty}\phi(\bz^{\l_{n}})\,.
\ee
Irreducible representations of the symmetric group $S_{N}$ are also labeled by a Young tableau. If $\sigma\in S_{N}$ and $X^{\l}(\sigma)$ is the associated matrix, we define
$\chi^{\l}(\sigma) = \Tr X^{\l}(\sigma)$.
Conjugacy classes in $S_{N}$ are also labeled by a Young tableau, and corresponds to the cycle structure of a class representative. Let they be $K_{\mu}$. We define
$\wh\chi^{\l}_{\mu}  = \chi^{\l}(\sigma)$ with $\sigma\in K_{\mu}$. These object can be computed by the Murnaghan-Nakayama rule, see for instance \cite{stanley2023enumerative}.
One has the two completeness relations
\bea
& \frac{1}{N!}\sum_{\sigma\in S_{N}}\chi^{\l}(\sigma)\chi^{\mu}(\sigma) = \sum_{|\nu|=N}\frac{1}{\z_{\nu}}\wh\chi^{\l}_{\nu}\wh\chi^{\mu}_{\nu} = \delta_{\l \mu}, \\
& \sum_{|\nu|=N}\wh\chi^{\nu}_{\l}\wh\chi^{\nu}_{\mu} = \z_{\l}\, \delta_{\l \mu}\,,
\eea
and the important formula \cite{stanley2023enumerative}
\be
\langle \chi_{\smb}^{\l}, \chi_{\smb}^{\mu}\rangle_{N}  = \delta_{|\l|,|\mu|}\mathop{\sum_{|\nu|=|\l|}}_{\ell(\nu)\le N}\wh\chi^{\nu}_{\l}\wh\chi^{\nu}_{\mu}\,.
\ee
Hence, (\ref{2.5}) can be written as 
\be
\la{2.9}
Z_{N}(\bg) = \sum_{d=0}^{\infty}\sum_{|\l|=d}\frac{\bg^{\l}}{\z_{\l}}\ \mathop{\sum_{|\nu|=d}}_{\ell(\nu)\le N}(\wh\chi^{\nu}_{\l})^{2}\,.
\ee
The Schur line defect 2-point function in the fundamental $\I^{U(N)}_{\rm F}(\eta; q)$ is obtained from an object similar to (\ref{2.5}), \ie
\be
Z_{N}^{\rm F}(\bg) = \sum_{d=0}^{\infty}\sum_{|\l|=d}\frac{\bg^{\l}}{\z_{\l}}\ \langle \chi_{\smb}(\bz)
\chi_{\smb}^{\l}(\bz), \chi_{\smb}(\bz^{-1})\chi_{\smb}^{\l}(\bz)\rangle_{N}\,.
\ee
Let us denote by $\l'$ the Young Tableaux $\l$ with the addition of one block at the bottom, so with one more row of length 1. Obviously
\be
\chi_{\smb}(\bz) \chi_{\smb}^{\l}(\bz) = \chi_{\smb}^{\l'}(\bz), \qquad \z_{\l'} = (r_{1}+1)\, \z_{\l}\,.
\ee
and thus
\ba
\la{2.12}
Z_{N}^{\rm F}(\bg) &=
\sum_{d=0}^{\infty}\sum_{|\l|=d}\frac{r_{1}+1}{\z_{\l}'}\bg^{\l}\ \mathop{\sum_{|\nu|=d+1}}_{\ell(\nu)\le N}(\wh\chi^{\nu}_{\l'})^{2}\,. 
\ea
The large $N$ limits discussed previouly are clearly reproduced by these expressions. For $Z_{N}(\bg)$ one has indeed \cite{Murthy:2022ien}
\be
Z_{\infty}(\bg) = \sum_{\l}\bg^{\l} = \prod_{n=1}^{\infty}\sum_{r_{n}=0}^{\infty}g_{n}^{r_{n}} = \prod_{n=1}^{\infty}\frac{1}{1-g_{n}}\,,
\ee
and this is easily generalized to the 2-point function of the fundamental representation (or other cases) 
\ba
\la{2.14}
Z_{\infty}^{\rm F}(\bg)=\sum_{\l}(r_{1}+1) \bg^{\l} = \sum_{r_{1}=0}^{\infty}(r_{1}+1)g_{1}^{r_{1}}\times \prod_{n=2}^{\infty}
\sum_{r_{n}=0}^{\infty}g_{n}^{r_{n}} = \frac{1}{1-g_{1}}\prod_{n=1}^{\infty}\frac{1}{1-g_{n}}\,.
\ea
To go beyond the large $N$ limit, we start from  the representation (\ref{2.12}) for the  index, \cf (\ref{22}),
\be
\I^{U(N)}_{\rm F}(\eta; q) =
\sum_{d=0}^{\infty}\sum_{|\l|=d}\frac{r_{1}+1}{\z_{\l}'}f^{\l}(\eta; q)\ \mathop{\sum_{|\nu|=d+1}}_{\ell(\nu)\le N}(\wh\chi^{\nu}_{\l'})^{2},
\ee
with, \cf (\ref{2.6}),
\be
f^{\l}(\eta; q) = \prod_{n=1}^{\infty}f(\eta^{\l_{n}}; q^{\l_{n}})\,.
\ee
The first correction with respect to the $N\to \infty$ limit (\ref{2.14})
is due to Young tableaux with  $|\l|=N$. Recall also that $f^{\l}(\eta; q) = \mc O(q^{|\l|})$. We have thus
\ba
\I_{\rm F}^{U(N)}-\I_{\rm F}^{U(\infty)} &= \sum_{d=N}^{\infty}\sum_{|\l|=d}\frac{r_{1}+1}{\z_{\l'}}f_{\l}(\eta; q)\ \bigg[
\mathop{\sum_{|\nu|=d+1}}_{\ell(\nu)\le N}(\wh\chi^{\nu}_{\l'})^{2}-\sum_{|\nu|=d+1}(\wh\chi^{\nu}_{\l'})^{2}\bigg] 
=  \sum_{d=N}^{\infty} J_{d},
\ea
with 
\ba
J_{d} = -\sum_{|\l|=d}\frac{r_{1}+1}{\z_{\l'}}f_{\l}(\eta; q)\ 
\mathop{\sum_{|\nu|=d+1}}_{\ell(\nu)\ge N+1}(\wh\chi^{\nu}_{\l'})^{2}.
\ea
The leading term has $d=N$ and therefore corresponds to the unique partition  
$\nu = (1, \dots, 1) = (1^{N+1})$ for which $(\wh\chi^{\nu}_{\l'})^{2}=1$ \cite{Dolan:2007rq}.
Thus
\be
J_{N} = -\sum_{|\l|=N}\frac{1}{\z_{\l}}f_{\l}(\eta; q)  = -\PE[\eps f(\eta; q)]\big|_{\eps^{N}}, \qquad \PE[\eps f(\eta; q)] = \exp\bigg[\sum_{n=1}^{\infty}
\frac{\eps^{n}}{n}f(\eta^{n}; q^{n})\bigg].
\ee
In our specific case, we have 
\be
f(\eta; q) = (\eta+\eta^{-1})\, q+\mc O(q^{2})\, ,
\ee
and therefore
\ba
\frac{\I^{U(N)}(\eta; q)}{\I^{U(\infty)}(\eta; q)} &= 1-\PE[\eps (\eta+\eta^{-1})]\big|_{\eps^{N}}\, q^{N}+\mc O(q^{N+1}).
\ea
The coefficient $\eps^{N}$ in the plethystic is easily computed from 
\ba
\la{3.15}
-\PE[\eps(\eta+\eta^{-1})]\big|_{\eps^{N}} = -\oint_{|z|=r}\frac{dz}{2\pi i z^{N+1}}\frac{1}{1-\eps\eta}\frac{1}{1-\eps\eta^{-1}} =  \frac{\eta}{1-\eta^{2}}(\eta^{N+1}-\eta^{-N-1}),
\ea
where the circle radius $r$ is taken small enough to exclude the poles $\eps = \eta^{\pm 1}$ and the integral is then done by picking up residues. The above 
is in agreement with the finite $N$ analysis, \cf (\ref{x38}).

\subsection{Exact single giant graviton correction $G^{+}_{\texorpdfstring{\rm F}{F}}(\eta; q)$}

Let us now compute the exact sum of all missing contributions in (\ref{x38}), \ie the exact function $G^{+}_{\rm F}(\eta; q)$ in (\ref{3.4})
whose first terms in the small $q$ expansion were given in (\ref{3.5}).
We consider again the multi-coupling matrix integral $Z_{N}(\bg)$ introduced in (\ref{2.1}). The leading single giant graviton correction 
has been worked out in general in \cite{Murthy:2022ien} and reads 
\be
\frac{Z_{N}(\bg)}{Z_{\infty}(\bg)} = 1+G_{N}(\bg)+\cdots\,,
\ee
with 
\be
\la{3.17}
G_{N}(\bg) = -\frac{\zeta}{(1-\zeta)^{2}}\exp\bigg[-\sum_{n=1}^{\infty}\frac{1}{n}\frac{g_{n}}{1-g_{n}}(1-\zeta^{n})(1-\zeta^{-n})\bigg]\bigg|_{\zeta^{-N}}.
\ee
This formula is the first contribution from a determinantal expansion and is expected to be valid up to terms where two gravitons contributions are important, 
\ie generically $\sim q^{2N}$, see footnote \ref{foot1}.
The full higher order giant graviton expansion was discussed in \cite{Murthy:2022ien}
and interpreted as an instanton expansion in \cite{Eniceicu:2023cxn}. The two giant graviton contribution was worked out in \cite{Liu:2022olj}. These works examined the 
unitary matrix integral representation of the index. A discussion of the comparison with wrapped brane expansion beyond the single giant graviton contribution 
was presented in \cite{Eniceicu:2023uvd}.

For the Schur line defect 2-point function in fundamental representation, we need 
\be
Z_{N}^{\rm F}(\bg) = \int_{U(N)}dU\, \Tr U\Tr U^{-1}\exp\bigg(\sum_{n=1}^{\infty}\frac{1}{n}g_{n}\Tr U^{n}\Tr U^{-n}\bigg)\,,
\ee
and we recall relations (\ref{x24}). 
The giant graviton expansion of $Z_{N}^{\rm F}(\bg)$ is then given by 
\ba
\frac{Z_{N}^{\rm F}(\bg)}{Z_{\infty}^{\rm F}(\bg)} &= \frac{\partial_{g_{1}}Z_{N}(\bg)}{\frac{1}{1-g_{1}}Z_{\infty}(\bg)}
= (1-g_{1})\frac{1}{Z_{\infty}(\bg)}\partial_{g_{1}}[Z_{\infty}(\bg)(1+G_{N}(\bg)+\cdots)] \lp
= (1-g_{1})\partial_{g_{1}}\log Z_{\infty}(\bg)\, (1+G_{N}(\bg)+\cdots)+(1-g_{1})\partial_{g_{1}}G_{N}(\bg)+\cdots \lp
= 1+G_{N}(\bg)+(1-g_{1})\partial_{g_{1}}G_{N}(\bg)+\cdots\,.
%= 1+(1-g_{1})^{2}\partial_{g_{1}}\bigg[\frac{1}{1-g_{1}}G_{N}(\bg)\bigg]+\cdots\,.
\ea
Thus, we may write 
\be
\frac{Z_{N}^{\rm F}(\bg)}{Z_{\infty}^{\rm F}(\bg)} = 1+G_{N}^{\rm F}(\bg)+\cdots, \qquad 
G_{N}^{\rm F}(\bg) = G_{N}(\bg)-\frac{1}{1-g_{1}}\,E_{N}(\bg)\,,
\ee
where, using (\ref{3.17}), we find
\ba
E_{N}(\bg) &= \exp\bigg[-\sum_{n=1}^{\infty}\frac{1}{n}\frac{g_{n}}{1-g_{n}}(1-\zeta^{n})(1-\zeta^{-n})\bigg]\bigg|_{\zeta^{-N}}.
\ea
This is the same plethystic as in (\ref{3.17}) up to the missing prefactor $-\zeta/(1-\zeta)^{2}$ and thus can be evaluated  as a finite difference of $G_{N}(\bg)$ functions.
To streamline notation, let us denote
\be
H(\bg, \zeta) = \frac{\bg}{1-\bg}(1-\zeta)(1-\zeta^{-1}), \qquad G(\bg, \zeta) = \PE[-H(\bg, \zeta)], \qquad G_{N}(\bg) = G(\bg, \zeta)|_{\zeta^{-N}}\, .
\ee
We have simply
\ba
E_{N}(\bg) &= \oint\frac{d\zeta}{\zeta^{N+1}} \PE[-H(\bg, \zeta)] = 
-\oint\frac{d\zeta}{\zeta^{N+1}} \frac{(1-\zeta)^{2}}{\zeta}\,G(\bg, \zeta) = -G_{N+1}(\bg)+2G_{N}(\bg)-G_{N-1}(\bg)\,,
\ea
which is valid for $N\ge 2$. This means that we have obtained the single giant graviton correction to $\I_{\rm F}^{U(N)}(\eta; q)$ in the form
\ba
\frac{\I^{U(N)}_{\rm F}(\eta; q)}{\I^{U(\infty)}_{\rm F}(\eta; q)} = 1+G_{N}^{\rm F}(\eta; q) + \cdots\, ,
\ea
with the following finite-difference expression
\ba
\la{3.27}
G_{N}^{\rm F}(\eta; q) &= G_{N}(\eta; q)+\frac{1}{1-f(\eta; q)}\,[G_{N+1}(\eta; q)-2G_{N}(\eta; q)+G_{N-1}(\eta; q)]\, .
\ea
From (\ref{3.1}), the exact expression of $G_{N}(\eta; q)$ is
\be
\la{3.28}
G_{N}(\eta; q) = -q^{N+1}\,\bigg[\eta^{N+2}\frac{(\frac{q}{\eta})_{\infty}^{3}}{\vth(\eta^{2},\frac{q}{\eta})}
+\eta^{-N-2}\frac{(\eta\,q)_{\infty}^{3}}{\vth(\eta^{-2},\eta\,q)}\bigg],
\ee
and this gives
\ba
& G_{N}^{\rm F}(\eta; q) = \lp
 -q^{N+1}\,\bigg[\eta^{N+2}\bigg(
1+\frac{1-q^{2}}{\eta\, q}\frac{1-\eta\, q}{1-\eta^{-1}\, q}
\bigg)\frac{(\frac{q}{\eta})_{\infty}^{3}}{\vth(\eta^{2},\frac{q}{\eta})}+
\eta^{-N-2}\bigg(
1+\frac{1-q^{2}}{\eta^{-1}\, q}\frac{1-\eta^{-1}\, q}{1-\eta\, q}
\bigg)\frac{(\eta\,q)_{\infty}^{3}}{\vth(\eta^{-2},\eta\,q)}
\bigg].
\ea
Comparing with (\ref{3.4}), the exact expression of the function $G^{+}_{\rm F}$ is thus
\be
\la{x341}
G^{+}_{\rm F}(\eta; q) = -\eta^{2}\,q\,\bigg(
1+\frac{1-q^{2}}{\eta\, q}\frac{1-\eta\, q}{1-\eta^{-1}\, q}
\bigg)\frac{(\frac{q}{\eta})_{\infty}^{3}}{\vth(\eta^{2},\frac{q}{\eta})}\, .
\ee
As a check, one can expand it in small $q$ and reproduce (\ref{3.5}). In the unflavored limit we have, \cf (\ref{3.6}), 
\be
G_{N}(1; q) = -(N+2)\, q^{N+1}.
\ee
Using (\ref{3.27}) gives then 
\be
G_{N}^{\rm F}(1; q) = -\bigg[N+1+N\,q-(N+3)\,q^{2}\bigg]\, q^{N}\, ,
\ee
which is exact and shows that (\ref{3.7}) has actually three terms without further corrections.
As a final comment, it is clear that the factorization in (\ref{x341}) is a simple consequence of the finite-difference structure in (\ref{3.27}).

\subsection{Schur line defect correlators of charged Wilson lines}

Generalization to the case of insertion of a pair of Wilson lines with opposite charges $\pm Q$  is straightforward by the same method. 
Now, we differentiate with respect to $g_{Q}$ and start from 
\be
Z_{N}^{Q}(\bg) = Q\frac{\partial}{\partial g_{Q}}Z_{N}(\bg)\,.
\ee
For $N=\infty$, we get \footnote{Our formalism provides immediately the large $N$ limit of Schur defect line correlators with general insertions 
proving easily the conjectures in Section 5.1.1 of \cite{Hatsuda:2023iwi}.}
\be
Z_{\infty}^{Q}(\bg) = Q\frac{\partial}{\partial g_{Q}}\prod_{n=1}^{\infty}\frac{1}{1-g_{n}} = \frac{Q}{1-g_{Q}}\prod_{n=1}^{\infty}\frac{1}{1-g_{n}}\, .
\ee
The single giant graviton correction is given by 
\ba
\frac{Z_{N}^{Q}(\bg)}{Z_{\infty}^{Q}(\bg)} &= \frac{Q\partial_{g_{Q}}Z_{N}(\bg)}{\frac{Q}{1-g_{Q}}Z_{\infty}(\bg)}
= (1-g_{Q})\frac{1}{Z_{\infty}(\bg)}\partial_{g_{Q}}[Z_{\infty}(\bg)(1+G_{N}(\bg)+\cdots)] \lp
= 1+G_{N}(\bg)+(1-g_{Q})\partial_{g_{Q}}G_{N}(\bg)+\cdots\, .
\ea
We now observe that 
\ba
(1-g_{Q})& \partial_{g_{Q}}G_{N}(\bg) = -\frac{1}{Q}\frac{1}{1-g_{Q}}\frac{\zeta^{1-Q}(1-\zeta^{Q})^{2}}{(1-\zeta)^{2}}\PE[-H(\bg, \zeta)]\lp
= \frac{1}{Q}\frac{1}{1-g_{Q}}\zeta^{-Q}(1-\zeta^{Q})^{2}\bigg[-\frac{\zeta}{(1-\zeta)^{2}}\PE[-H(\bg, \zeta)]\bigg]\, .
\ea
By the same manipulations as in the $Q=1$ case, it  follows
\be
(1-g_{Q})\partial_{g_{Q}}G_{N}(\bg) =\frac{1}{Q} \frac{1}{1-g_{Q}}[G_{N+Q}(\bg)-2G_{N}(\bg)+G_{N-Q}(\bg)]\, ,
\ee
and thus
\ba
\frac{\I^{U(N)}_{Q}(\eta; q)}{\I^{U(\infty)}_{Q}(\eta; q)} = 1+G_{N}^{Q}(\eta; q) + \cdots\, ,
\ea
with 
\ba
G_{N}^{Q}(\eta; q) &= G_{N}(\eta; q)+\frac{1}{Q}\frac{1}{1-f(\eta^{Q}; q^{Q})}\,[G_{N+Q}(\eta; q)-2G_{N}(\eta; q)+G_{N-Q}(\eta; q)]\,.
\ea
Using again (\ref{3.28}), we obtain 
\ba
\la{48}
& G_{N}^{Q}(\eta; q) = \lp
 -q^{N+1}\,\bigg[\eta^{N+2}\bigg(
1+\frac{1}{Q}\frac{1-q^{2Q}}{\eta^{Q}\, q^{Q}}\frac{1-\eta^{Q}\, q^{Q}}{1-\eta^{-Q}\, q^{Q}}
\bigg)\frac{(\frac{q}{\eta})_{\infty}^{3}}{\vth(\eta^{2},\frac{q}{\eta})}+
\eta^{-N-2}\bigg(
1+\frac{1}{Q}\frac{1-q^{2Q}}{\eta^{-Q}\, q^{Q}}\frac{1-\eta^{-Q}\, q^{Q}}{1-\eta^{Q}\, q^{Q}}
\bigg)\frac{(\eta\,q)_{\infty}^{3}}{\vth(\eta^{-2},\eta\,q)}
\bigg].
\ea
The first correction appears now at order $q^{N+1-Q}$. In the unflavored limit $\eta\to 1$ it is exactly
\be
G_{N}^{Q}(1; q) = -q^{N+1-Q}\, \bigg[\frac{N+2}{Q}-1+N\,q^{Q}-\bigg(\frac{N+2}{Q}+1\bigg)\, q^{2Q}\bigg].
\ee

\section{The  $(\Box, \Box, \overline{\Box}, \overline{\Box})$ 4-point function}
\la{sec:four}

The previous methods can be clearly applied to more complicated cases. In particular we can consider the 4-point function with insertion of two lines
in the fundamental and two in the anti-fundamental. In this case, we need 
\be
Z_{N}^{\rm FF}(\bg) = \int_{U(N)}dU\, (\Tr U\Tr U^{-1})^{2}\exp\bigg(\sum_{n=1}^{\infty}\frac{1}{n}g_{n}\Tr U^{n}\Tr U^{-n}\bigg) = 
 \frac{\partial^{2}}{\partial g_{1}^{2}}Z_{N}(\bg)\,.
\ee
The large $N$ limit is 
\be
\la{52}
Z_{\infty}^{\rm FF}(\bm{g}) = \frac{2}{(1-g_{1})^{2}}\prod_{n=1}^{\infty}\frac{1}{1-g_{n}}\, .
\ee
The single giant graviton correction is computed from 
\ba
\frac{Z_{N}^{\rm FF}(\bg)}{Z_{\infty}^{\rm FF}(\bg)} &= \frac{\partial_{g_{1}}^{2}Z_{N}(\bg)}{\frac{2}{(1-g_{1})^{2}}Z_{\infty}(\bg)}
= \frac{1}{2}(1-g_{1})^{2}\frac{1}{Z_{\infty}(\bg)}\partial_{g_{1}}^{2}[Z_{\infty}(\bg)(1+G_{N}(\bg)+\cdots)] \lp
= 1+G_{N}(\bg)+(1-g_{1})\partial_{g_{1}}G_{N}(\bg)+\frac{1}{2}(1-g_{1})^{2}\partial_{g_{1}}^{2}G_{N}(\bg)+\cdots \, .
\ea
This gives 
\be
\frac{Z_{N}^{\rm FF}(\bg)}{Z_{\infty}^{\rm FF}(\bg)} = 1+G_{N}^{\rm FF}(\bg)+\cdots\, ,
\ee
with 
\ba
& G_{N}^{\rm FF}(\bg) = G_{N}(\bg)-\bigg[\frac{1-2g_{1}}{(1-g_{1})^{2}}+\frac{1}{(1-g_{1})^{2}}\frac{1+\zeta^{2}}{2\zeta}\bigg]\, \PE[-H(\bg, \zeta)]\lp
= G_{N}(\bg)+\bigg[\frac{(1-\zeta)^{2}}{\zeta}\frac{1-2g_{1}}{(1-g_{1})^{2}}+\frac{1}{(1-g_{1})^{2}}\frac{(1+\zeta^{2})(1-\zeta)^{2}}{2\zeta^{2}}\bigg]\, \frac{-\zeta}{(1-\zeta)^{2}}\PE[-H(\bg, \zeta)]\, .
\ea
For the index, this implies
\ba
\frac{\I^{U(N)}_{\rm FF}(\eta; q)}{\I^{U(\infty)}_{\rm FF}(\eta; q)} = 1+G_{N}^{\rm FF}(\eta; q) + \cdots\,  = 1+[\eta^{N}G_{\rm FF}^{+}(\eta; q)
+\eta^{-N}G_{\rm FF}^{-}(\eta; q)]q^{N}+\cdots
\ea
with the following exact finite-difference expression ($f \equiv f(\eta, q), G_{N}\equiv G_{N}(\eta; q)$)
\ba
\la{x47}
G_{N}^{\rm FF}(\eta; q) &= \frac{f\,(2+f)}{(1-f)^{2}}G_{N}+\frac{1}{2(1-f)^{2}}(G_{N+2}+G_{N-2})-\frac{2f}{(1-f)^{2}}(G_{N+1}+G_{N-1})\, .
\ea
Using (\ref{3.28}), we get 
\ba
\la{x48}
G_{\rm FF}^{+}(\eta; q) =-\eta^{2}\, q\, \bigg[
1+\frac{1-q^{2}}{2\eta^{2}\, q^{2}}\frac{1-\eta\, q}{(1-\eta^{-1}\, q)^{2}}(1-5q^{2}+3\eta q+\eta q^{3})
\bigg]\frac{(\frac{q}{\eta})_{\infty}^{3}}{\vth(\eta^{2},\frac{q}{\eta})}\, ,
\ea
with explicit expansion 
\ba
G_{\rm FF}^{+}(\eta; q) =
\frac{\eta }{2\,(1- \eta ^2) }\frac{1}{q}+\frac{1+3 \eta ^4}{2 \eta ^2(1-\eta^{2})}+\bigg(\frac{1}{2 \eta ^5}+\frac{1}{\eta ^3}+\frac{3}{2 \eta }-\eta \bigg)\,q+\cdots\, .
\ea
This expression reproduces the matrix integral series up to the order $q^{2N}$ where double giant graviton contributions appear. In the unflavored limit $\eta\to 1$, we get 
the polynomial correction
\be
G_{N}^{\rm FF}(1; q) = -[N+4N\, q-8q^{2}-4\, (N+2)\, q^{3}+(N+4)\, q^{4}]\, q^{N-1}\, .
\ee
Again, the factorized structure of (\ref{x48}) is a direct consequence of the finite-difference representation (\ref{x47}).

\section{Single giant graviton correction from string fluctuations}
\la{sec:gravity}

For the Schur index without insertions, the giant graviton expansion (\ref{3.1}) may be written in the form (\ref{3.4}) as 
\be
\frac{\I^{U(N)}(\eta; q)}{\I^{U(\infty)}(\eta; q)} = 
1 +\bigg[\eta^{N}G^{+}_{\rm D3}(\eta; q)+\eta^{-N}G^{-}_{\rm D3}(\eta; q)\bigg]\, q^{N}+\mc O(q^{2N})\, ,
\ee
where  
\be
G^{+}_{\rm D3}(\eta; q) = -\eta^{2}q\, \frac{(\frac{q}{\eta})_{\infty}^{3}}{\vth(\eta^{2},\frac{q}{\eta})} = 
\PE\bigg[\frac{\frac{1}{\eta q}-\frac{2}{\eta}q+q^{2}}{1-\frac{q}{\eta}}\bigg]\, ,
\ee
is the single giant graviton contribution from wrapped D3 brane \cite{Arai:2020qaj,Beccaria:2024szi}.

\paragraph{The 2-point function with representations $(\Box, \overline{\Box})$}

As we mentioned in the introduction, the large $N$ limit of $\I_{\rm F}^{U(\infty)}(\eta; q)$ 
may be written in the factorized  form 
\be
\I_{\rm F}^{U(\infty)}(\eta; q) = \I_{\rm F1}(\eta; q)\,\, \I^{U(\infty)}(\eta; q), 
\ee
where
\be
I_{\rm F1}(\eta; q) = \frac{1}{1-f(\eta; q)}=  \frac{1-q^{2}}{(1-\eta q)(1-\eta^{-1}q)} = \PE[-q^{2}+(\eta+\eta^{-1})\, q]\, ,
\ee
is the index of fluctuations of a fundamental string along $AdS_{2}\subset AdS_{5}$. Indeed the simple three term argument of the plethystic is the associated single particle index 
as shown in \cite{Gang:2012yr}. 

As suggested in \cite{Imamura:2024lkw}, in the analysis of finite $N$ corrections to $\I_{\rm F}^{U(N)}$, one should consider two possibilities for the 
coexistence of the string worldsheet and a single giant graviton. The first case is when it does not end on the giant world-volume of the giant. The second is
when the string world-sheet is separated by the giant and the two semi-infinite strings end on it. Subtraction of the contribution from the first case 
isolates the latter possibility. This leads to 
consider the ratio 
\ba
\la{65}
\frac{\I_{\rm F}^{U(N)}-I_{\rm F1}\,\I^{U(N)}}{\I^{U(\infty)}} &= I_{\rm F1}\left(\frac{\I_{\rm F}^{U(N)}}{\I_{\rm F}^{U(\infty)}} -
\frac{\I^{U(N)}}{\I^{U(\infty)}}\right)\, ,
\ea
where the difference has been divided by the supergravity contribution in absence of the defect, \ie we do not divide by $\I_{\rm F}^{U(\infty)}$ as we did so far.
This gives
\be
\frac{\I_{\rm F}^{U(N)}-I_{\rm F1}\,\I^{U(N)}}{\I^{U(\infty)}}  = 1+\bigg(\mc G^{+}_{\rm F}(\eta; q)\, \eta^{N}+\mc G^{-}_{\rm F}(\eta; q)\,\eta^{-N}\bigg)\, q^{N}+\mc O(q^{2N})\,, 
\ee
with 
\ba
\la{4.5}
\mc G^{+}_{\rm F}  &= I_{\rm F1}(G_{\rm F}^{+}-G^{+}_{\rm D3}) = \frac{1}{\eta q}\frac{(1-q^{2})^{2}}{(1-\eta^{-1}q)^{2}}
\PE\bigg[\frac{\frac{1}{\eta q}-\frac{2}{\eta}q+q^{2}}{1-\frac{q}{\eta}}\bigg] \lp
= \frac{1}{\eta q}\PE\bigg[\frac{\frac{1}{\eta q}-\frac{2}{\eta}q+q^{2}}{1-\frac{q}{\eta}}+2\eta^{-1}q-2q^{2} \bigg]\, .
\ea
In the conventions of \cite{Imamura:2024lkw} we have $q = \sqrt{xy}$ and $\eta = \sqrt{x/y}$. The extra single particle index in (\ref{4.5}) is thus
\be
\la{58}
2\eta^{-1}q-2q^{2} = 2\,(1-x)\,y\, ,
\ee
in agreement with the analysis in \cite{Imamura:2024lkw} of the fluctuation modes on a  world-sheet  along $AdS_{2}\subset AdS_{5}$  studied in \cite{Drukker:2000ep,Faraggi:2011bb}.
In more details, without defect lines, there are five bosonic scalar fluctuations in the fundamental of $SO(5)$ corresponding to $S^{5}$ coordinates. 
Together with the three scalar fluctuations in the remaining coordinates of $AdS_{5}$ plus fermionic states, these are part of a supermultiplet with $8_{\rm B}+8_{\rm F}$ 
states with  $SO(1,2)\times SO(3)\times SO(5)$ quantum numbers 
\be
\text{B}:\ (1,0,\bm{5})\oplus(2,1,\bm{1}), \qquad \text{F}:\ (\tfrac{3}{2}, \tfrac{1}{2}, \bf{4})\, .
\ee
The defect lines break it to $SO(1,2)\times SO(2)\times SO(3)$ 
and the five scalars split into $\phi_{3}$ in the triplet of $SO(3)$ plus two scalars $\varphi_{i}$, $i=1,2$ in the singlet. BPS states contributing the index
are three, \ie one component of $\phi_{3}$,
one of the two scalars $\varphi_{i}$ and one fermionic state. Dirichlet boundary conditions on the giant graviton remove one of the scalars
 and leave two BPS states corresponding
to the two contributions in (\ref{58}).

Our derivation reproduces the peculiar prefactor $1/(\eta q)$ in (\ref{4.5}), whose origin 
is at the moment unclear from the point of view of string fluctuations and has been suggested to be related 
to a back-reaction of the fundamental strings in \cite{Imamura:2024lkw}.

When the pair of Wilson lines have charges $Q, -Q$, the expression (\ref{4.5}) has to be simply changed to 
\ba
\la{4.5}
\mc G^{+}_{Q}  &= G_{\rm D3}^{+}(\eta; q) \times \frac{1}{(\eta q)^{Q}}\PE\bigg[2\eta^{-Q}q^{Q}-2q^{2Q} \bigg]\, ,
\ea
as follows from (\ref{48}).

\paragraph{The 4-point function with representations $(\Box, \Box, \overline{\Box}, \overline{\Box})$}

It is interesting to examine what we get in the case of the 4-point function with two fundamental lines and two anti-fundamental lines. The large $N$ limit 
of the index is $2\I_{\rm F1}^{2}$, \cf (\ref{52}). So, the natural generalization of (\ref{65}) reads
\ba
\la{610}
\frac{\I_{\rm FF}^{U(N)}-2I_{\rm F1}^{2}\,\I^{U(N)}}{\I^{U(\infty)}} &= 2I_{\rm F1}^{2}\left(\frac{\I_{\rm FF}^{U(N)}}{\I_{\rm FF}^{U(\infty)}} -
\frac{\I^{U(N)}}{\I^{U(\infty)}}\right)\, ,
\ea
and its expansion takes the form 
\be
\frac{\I_{\rm FF}^{U(N)}-2I_{\rm F1}^{2}\,\I^{U(N)}}{\I^{U(\infty)}}  = 1+\bigg(\mc G^{+}_{\rm FF}(\eta; q)\, \eta^{N}+\mc G^{-}_{\rm FF}(\eta; q)\,\eta^{-N}\bigg)\, q^{N}+\mc O(q^{2N})\,, 
\ee
with 
\ba
\la{4.5}
\mc G^{+}_{\rm FF}  &= 2I_{\rm F1}^{2}(G_{\rm FF}^{+}-G^{+}_{\rm D3}) = \frac{1-5q^{2}+3\eta q+\eta q^{3}}{\eta^{2} q^{2}}\frac{(1-q^{2})^{3}}{(1-\eta^{-1}q)^{4}(1-\eta q)}
\PE\bigg[\frac{\frac{1}{\eta q}-\frac{2}{\eta}q+q^{2}}{1-\frac{q}{\eta}}\bigg] \lp
= \frac{1+3\eta q-5q^{2}+\eta q^{3}}{\eta^{2} q^{2}}\PE\bigg[\frac{\frac{1}{\eta q}-\frac{2}{\eta}q+q^{2}}{1-\frac{q}{\eta}}-3q^{2}+4\eta^{-1}q+\eta q \bigg]\, .
\ea
Comparing with (\ref{4.5}), we have now a more complicated prefactor which however is still a sum of monomials. The extra contribution in the plethystic exponential
should come from fluctuations of a  world-sheet attached to the four lines and ending on the giant graviton. Here, the geometry is  more complicated
and additional subtractions could be needed in (\ref{610}) to simplify the result. 
Still, it seems clear that a better deeper understanding of the prefactor origin is definitely worth.

\section*{Acknowledgements}

We thank Arkady Tseytlin, Ji Hoon Lee, and Alejandro Cabo-Bizet for useful discussions related to various aspects of this work. 
Financial support from the INFN grant GAST is acknowledged.

\appendix
\section{Special functions}
\la{app:special}

We collect in this appendix the definition of  special functions appearing in the text.

\paragraph{$q$-Pochhammer symbol}

\ba
\la{A.1}
(a; q)_{\infty}&= \prod_{k=0}^{\infty}(1-a\,q^{k})\,, \qquad (a^{\pm}; q)_{\infty} = (a; q)_{\infty}(a^{-1};  q)_{\infty}\, , \\
(q)_{\infty} &\equiv (q; q)_{\infty} = \prod_{k=1}^{\infty}(1-q^{k})\, .
\ea

\paragraph{$q$-theta function}

The $q$-theta function is defined as 
\be
\la{A.3}
\vth(x,q) = -x^{-\frac{1}{2}}(q)_{\infty}(x; q)_{\infty}(qx^{-1}; q)_{\infty}\, ,  
\ee
with 
\ba
\la{A.4}
\vth(x;q) = -\vth(x^{-1};q), \qquad \vth(q^{m}x; q) = (-1)^{m}q^{-\frac{m^{2}}{2}}x^{-m}\vth(x;q)\,.
\ea

\section{Correlators with general insertions and their single giant graviton correction}
\la{app:zero}

As we have seen, for general charge assignment of the Wilson lines, the large $N$ limit of the Schur defect correlator has a vanishing large $N$ limit. This happens
when the charge set $\bQ$ is not symmetric under a global change of sign of charges. 
In this case, $\I_{\bQ}^{U(N)}(\eta; q)$ has a small $q$ expansion that starts at order $q^{N}$, up to
an $N$ independent non-trivial function of the fugacities. The aim of this Appendix is to derive this function.

As discussed in \cite{Murthy:2022ien}, using free fermion methods, one computes the following first determinantal correction to $\wt Z_{N}(\bm{t}^{+}, \bm{t}^{-})$
in (\ref{x27})
\ba
\frac{\wt Z_{N}(\bm{t}^{+}, \bm{t}^{-})}{\wt Z_{\infty}(\bm{t}^{+}, \bm{t}^{-})} = 1-K_{N}(\bm{t}^{+}, \bm{t}^{-})+\cdots, \qquad
K_{N}(\bm{t}^{+}, \bm{t}^{-}) = \mathop{\sum_{N<r}}_{r\in\mathbb{Z}+\frac{1}{2}} \widetilde{K}(r,r; \bm{t}^{+}, \bm{t}^{-})\,,
\ea
with 
\bea
& \sum_{r,s\in\mathbb{Z}+\frac{1}{2}}\widetilde K(r,s; \bm{t}^{+}, \bm{t}^{-})\, z^{r}w^{-s} = \frac{J(z; \bm{t}^{+}, \bm{t}^{-})}{J(w; \bm{t}^{+}, \bm{t}^{-})}\, \frac{\sqrt{zw}}{z-w}, 
\qquad |w|<|z|,\\
& J(z; \bm{t}^{+}, \bm{t}^{-}) = \exp\bigg(\sum_{n=1}^{\infty}\frac{1}{n}(t_{n}^{+}z^{n}-t_{n}^{-}z^{-n})\bigg)\,. 
\eea
Including $\bQ$ charged insertions, we get 
\ba
D_{\bQ}\widetilde Z_{N}(\bm{t}^{+}, \bm{t}^{-}) &= D_{\bQ}[\widetilde Z_{\infty}(\bm{t}^{+}, \bm{t}^{-})-
\widetilde Z_{\infty}(\bm{t}^{+}, \bm{t}^{-})K_{N}(\bm{t}^{+}, \bm{t}^{-})+\cdots]\lp
= \widetilde Z_{\infty}^{\bQ}(\bm{t}^{+}, \bm{t}^{-})-D_{\bQ}[\widetilde Z_{\infty}(\bm{t}^{+}, \bm{t}^{-})K_{N}(\bm{t}^{+}, \bm{t}^{-})]+\cdots\, ,
\ea
where $D_{\bQ}$ was defined in (\ref{x211}).
Then, assuming $Z_{\infty}^{\bQ}(\bm g)=0$, the first term drops and we have
\ba
Z_{N}^{\bQ}(\bg) &= \langle D_{\bQ} \widetilde Z_{N}(\bm{t}^{+}, \bm{t}^{-})\rangle_{\bg}  = 
-\langle D_{\bQ}[\widetilde Z_{\infty}(\bm{t}^{+}, \bm{t}^{-})\, K_{N}(\bm{t}^{+}, \bm{t}^{-})]\rangle_{\bm g}+\cdots.
\ea
The right hand side can be evaluated by computing the r.h.s. of  
\ba
\la{x2}
& \sum_{r,s\in\mathbb{Z}+\frac{1}{2}}z^{r}w^{-s}\langle D_{\bQ}[\widetilde Z_{\infty}(\bm{t}^{+}, \bm{t}^{-})\, \widetilde K(r,s; \bm{t}^{+}, \bm{t}^{-})]\rangle_{\bm g} \lp
= \frac{\sqrt{zw}}{z-w}\int \prod_{k=1}\bigg[\frac{dt_{k}^{+} dt_{k}^{-}}{2\pi k g_{k}}\, \exp\bigg(-\frac{1}{kg_{k}}t_k^{+}t_{k}^{-}\bigg)\bigg]\ D_{\bQ}
\prod_{k=1}^{\infty}
\exp\frac{1}{k}\bigg(t_{k}^{+}t_{k}^{-}+t_{k}^{+}(z^{k}-w^{k})-t_{k}^{-}(z^{-k}-w^{-k})\bigg)\,,
\ea
which is a straightforward calculation given the precise charges  $\bQ$. For the purpose of illustration, let us examine 
 as a specific example the charge configuration $\bQ = (1,1; -2)$ or 
\be
D_{\bQ} = 2\partial_{t_{1}^{+}}^{2}\partial_{t_{2}^{-}}\,.
\ee
Integrating by parts gives
\be
-2\partial_{t_{1}^{+}}^{2}\partial_{t_{2}^{-}}\exp\bigg(-\sum_{k}\frac{1}{kg_{k}}t_k^{+}t_{k}^{-}\bigg) = \frac{t_{2}^{+}(t_{1}^{-})^{2}}{g_{1}^{2}g_{2}}\exp
\bigg(-\sum_{k}\frac{1}{kg_{k}}t_k^{+}t_{k}^{-}\bigg).
\ee
Thus,
\ba
& \sum_{r,s\in\mathbb{Z}+\frac{1}{2}}z^{r}w^{-s}\langle D_{(1,1; -2)}[\widetilde Z_{\infty}(\bm{t}^{+}, \bm{t}^{-})\, \widetilde K(r,s; \bm{t}^{+}, \bm{t}^{-})]\rangle_{\bm g} \lp
=\frac{1}{g_{1}^{2}g_{2}} \frac{\sqrt{zw}}{z-w}\int \prod_{k=1}\frac{dt_{k}^{+} dt_{k}^{-}}{2\pi k g_{k}}\,
t_{2}^{+}(t_{1}^{-})^{2}\, 
 \exp\frac{1}{k}\bigg(-\frac{1-g_{k}}{g_{k}}t_k^{+}t_{k}^{-}
+t_{k}^{+}(z^{k}-w^{k})-t_{k}^{-}(z^{-k}-w^{-k})\bigg).
\ea
We  have
\ba
& \int \prod_{k=1}\frac{dt_{k}^{+} dt_{k}^{-}}{2\pi k g_{k}}\
 \exp\frac{1}{k}\bigg(-\frac{1-g_{k}}{g_{k}}t_k^{+}t_{k}^{-}
+a_{k}t_{k}^{+}(z^{k}-w^{k})-b_{k}t_{k}^{-}(z^{-k}-w^{-k})\bigg) \lp
= \prod_{n=1}^{\infty}\frac{1}{1-g_{n}}\exp\bigg(-\sum_{k=1}^{\infty}\frac{1}{k}\frac{g_{k}}{1-g_{k}}a_{k}b_{k}(z^{k}-w^{k})(z^{-k}-w^{-k})\bigg)\lp
= \prod_{n=1}^{\infty}\frac{1}{1-g_{n}}\exp\bigg(\sum_{k=1}^{\infty}\frac{1}{k}\frac{g_{k}}{1-g_{k}}a_{k}b_{k}(-2+(w/z)^{-k}+(w/z)^{k})\bigg)\, , 
\ea
and we can bring down $t_{2}^{+}(t_{1}^{-})^{2}$ by applying to the above expression the operator
\be
\frac{2\cdot 1^{2}}{(z^{2}-w^{2})(z^{-1}-w^{-1})^{2}}\partial_{a_{2}}\partial_{b_{1}}^{2}\, , 
\ee
and setting $\bm{a}, \bm{b}\to \bm{1}$. We thus obtain (introducing $\gamma_{k}=g_{k}/(1-g_{k})$)
\ba
& \sum_{r,s\in\mathbb{Z}+\frac{1}{2}}z^{r}w^{-s}\langle D_{(1,1; -2)}[\widetilde Z_{\infty}(\bm{t}^{+}, \bm{t}^{-})\, \widetilde K(r,s; \bm{t}^{+}, \bm{t}^{-})]\rangle_{\bm g} \lp
=-\frac{1}{g_{1}^{2}g_{2}} \gamma_{1}^{2}\gamma_{2}\prod_{n=1}^{\infty}\frac{1}{1-g_{n}}\frac{(1-\frac{w}{z})^{3}(1+\frac{w}{z})}{(\frac{w}{z})^{2}}\, \frac{\sqrt{\frac{w}{z}}}{1-\frac{w}{z}}
\exp\bigg(\sum_{k=1}^{\infty}\frac{\gamma_{k}}{k}(-2+(w/z)^{-k}+(w/z)^{k})\bigg)\, .
\ea
The r.h.s. is a function of $w/z$ and thus the l.h.s. is  diagonal in $r,s$ and can be written
\ba
& \sum_{s\in\mathbb{Z}+\frac{1}{2}}\zeta^{-s}\langle D_{(1,1; -2)}[\widetilde Z_{\infty}(\bm{t}^{+}, \bm{t}^{-})\, \widetilde K(s,s; \bm{t}^{+}, \bm{t}^{-})]\rangle_{\bm g} \lp
=-\frac{1}{(1-g_{1})^{2}}\frac{1}{1-g_{2}}\prod_{n=1}^{\infty}\frac{1}{1-g_{n}}\frac{(1-\zeta)^{3}(1+\zeta)}{\zeta^{2}}\, \frac{\sqrt{\zeta}}{1-\zeta}
\exp\bigg(\sum_{k=1}^{\infty}\frac{\gamma_{k}}{k}(-2+\zeta^{k}+\zeta^{-k})\bigg).
\ea
Now, if we have 
\be
\sum_{s\in\mathbb{Z}+\frac{1}{2}}\zeta^{-s}H_{s} = f(\zeta)
\ee
we obtain  
\ba
\mathop{\sum_{N<s}}_{s\in\mathbb{Z}+\frac{1}{2}}H_{s} &= \mathop{\sum_{N<s}}_{s\in\mathbb{Z}+\frac{1}{2}}\oint d\zeta\, \zeta^{s-1}f(\zeta)
= \oint d\zeta f(\zeta)\, \sum_{n=0}^{\infty}\zeta^{N+\frac{1}{2}+n-1} = \int d\zeta \zeta^{N-1}\frac{\sqrt{\zeta}}{1-\zeta}f(\zeta) \lp
= \frac{\sqrt{\zeta}}{1-\zeta}f(\zeta)\bigg|_{\zeta^{-N}}.
\ea
This gives the final expression 
\ba
\la{6.27}
Z_{N}^{(1,1;-2)}(\bg) &= \bigg[\frac{1}{(1-g_{1})^{2}}\frac{1}{1-g_{2}}\prod_{n=1}^{\infty}\frac{1}{1-g_{n}}\bigg]\, \frac{1-\zeta^{2}}{\zeta}
\exp\bigg(\sum_{k=1}^{\infty}\frac{\gamma_{k}}{k}(-2+\zeta^{k}+\zeta^{-k})\bigg)\bigg|_{\zeta^{-N}}.
\ea
It is straightforward to write this in exact form as a finite sum over shifted $G_{N}(\bg)$. To test (\ref{6.27}), we  computed the exact matrix integral expansion 
of $\I_{(1,1;-2)}^{U(N)}(\eta; q)$ at finite $N=2,3,4, \dots$. With $s_{p} = \eta^{p}+\eta^{-p}$, the first three cases are 
\ba
I_{(1,1;-2)}^{U(2)}(\eta; q) &= s_{1}\,q+2 (s_{2}+1) \,q^2+(3s_{3}+s_{1})\,q^{3}
+(4s_{4}+2)\,q^4+\cdots, \notag \\
%%%%
\la{xbb}
I_{(1,1;-2)}^{U(3)}(\eta; q) &= (s_{2}+1)\,q^{2}+(2s_{3}+3s_{1})\, q^3+(4s_{4}+3s_{2}+4)\,q^4+(
6s_{5}+4s_{3}+4s_{1})\, q^{5}\lp
+(9s_{6}+3s_{4}+5s_{2}+4)\,q^6+\cdots, \\
%%%%
I_{(1,1;-2)}^{U(4)}(\eta; q) &= (s_{3}+s_{1})\, q^3+(2s_{4}+3s_{2}+4)\,q^4+
(4s_{5}+4s_{3}+6s_{1})\,q^5+(7s_{6}+6s_{4}+9s_{2}+6)\,q^6\lp
+(
11s_{7}+8s_{5}+10s_{3}+9s_{1})\, q^{7}+(
16s_{8}+9s_{6}+14s_{4}+7s_{2}+14)\,q^8+\cdots\, , \notag
\ea
and are fully reproduced by the perturbative series of (\ref{6.27}). In particular, the leading term is $\mc O(q^{N-1})$ and its coefficient is a function of $\eta$ that follows from  
\ba
\I_{(1,1;-2)}^{U(N)}(\eta; q) &= \PE[\eps \bm{\gamma}]|_{\eps^{N-1}} = \PE[\eps(\eta+\eta^{-1})]|_{\eps^{N-1}}\, q^{N-1}+\cdots\, .
\ea
Using (\ref{3.15}), we obtain 
\ba
\I_{(1,1;-2)}^{U(N)}(\eta; q) &= -\frac{\eta}{1-\eta^{2}}(\eta^{N}-\eta^{-N})\,q^{N-1}+\cdots = \mathop{\sum_{p=-(N-1)}^{N-1}}_{\Delta p = 2}\eta^{p}\ q^{N-1}+\cdots\, ,
\ea
in agreement with the corresponding terms in (\ref{xbb}). The procedure we have illustrated in this example can be easily applied to any other charge assignment.

\section{Explicit series expansion for the index $\I^{U(N)}_{\texorpdfstring{\rm F}{F}}(\eta; q)$}
\la{app:series}

We introduce  the notation $s_{p} = \eta^{p}+\eta^{-p}$. The explicit series expansion of the Schur line defect 2-point
function $\I_{\rm F}^{U(N)}$ at finite $N$ are  
\ba
\I^{U(2)}_{\rm F}(\eta; q) &= 1+2\,s_{1}\, q+(3\,s_{2}+1)\, q^{2}+4\,s_{3}\, q^{3}+\cdots
\ea
\ba
\I^{U(3)}_{\rm F}(\eta; q) &= 1+2\,s_{1}\, q+(4\,s_{2}+2)\, q^{2}+3\,(2\,s_{3}+s_{1})\, q^{3} 
+(9\,s_{4}+2\,s_{2}+3)\, q^{4}+\cdots
\ea
\ba
\I^{U(4)}_{\rm F}(\eta; q) &= 1+2\,s_{1}\, q+(4\,s_{2}+2)\, q^{2}+3\,(7\,s_{3}+4\,s_{1})\, q^{3} 
+(11\,s_{4}+5\,s_{2}+7)\, q^{4}\lp
+(16\,s_{5}+6\,s_{3}+7\,s_{1})\, q^{5}
+\cdots
\ea
\ba
\I^{U(5)}_{\rm F}(\eta; q) &= 1+2\,s_{1}\, q+(4\,s_{2}+2)\, q^{2}+3\,(7\,s_{3}+4\,s_{1})\, q^{3} 
+(12\,s_{4}+6\,s_{2}+8)\, q^{4}\lp
+(18\,s_{5}+9\,s_{3}+11\,s_{1})\, q^{5}
+(27\,s_{6}+10\,s_{4}+15\,s_{2}+13)\, q^{6}+\cdots
\ea
\ba
\I^{U(6)}_{\rm F}(\eta; q) &= 1+2\,s_{1}\, q+(4\,s_{2}+2)\, q^{2}+3\,(7\,s_{3}+4\,s_{1})\, q^{3} 
+(12\,s_{4}+6\,s_{2}+8)\, q^{4}\lp
+(19\,s_{5}+10\,s_{3}+12\,s_{1})\, q^{5}
+(29\,s_{6}+13\,s_{4}+19\,s_{2}+17)\, q^{6}\lp
+
(42\,s_{7}+18\,s_{5}+23\,s_{2}+25\,s_{1})\, q^{7}+\cdots
\ea
and 
\ba
\I^{U(\infty)}_{\rm F}(\eta; q) &= 1+2\,s_{1}\,q+(4\,s_{2}+2)\,q^{2}+(7\,s_{3}+4\,s_{1})\, q^{3}+(12\,s_{4}+6\, s_{2}+8)\, q^{4}\lp
+(19\,s_{5}+10\,s_{3}+12\,s_{1})\,q^{5}+(30\,s_{6}+14\,s_{4}+20\,s_{2}+18)\, q^{6}\lp
+(45\,s_{7}+22\,s_{5}+28\,s_{3}+30\,s_{1})\, q^{7}+\cdots\, .
\ea
We obtain 
\ba
\I^{U(2)}_{\rm F}-\I^{U(\infty)}_{\rm F} &= -\frac{1}{\eta^{2}}(1+\eta^{2}+\eta^{4})\, q^{2}-\frac{1}{\eta^{3}}(3+4\eta^{2}+4\eta^{4}+3\eta^{6})\, q^{3}+\cdots, \\
\I^{U(3)}_{\rm F}-\I^{U(\infty)}_{\rm F} &= -\frac{1}{\eta^{3}}(1+\eta^{2}+\eta^{4}+\eta^{6})\, q^{3}-\frac{1}{\eta^{4}}(3+4\eta^{2}+5\eta^{4}+4\eta^{6}+3\eta^{8})\, q^{4}+\cdots, \\
\I^{U(4)}_{\rm F}-\I^{U(\infty)}_{\rm F} &= -\frac{1}{\eta^{4}}(1+\eta^{2}+\eta^{4}+\eta^{6}+\eta^{8})\, q^{4}\lp
-\frac{1}{\eta^{5}}(3+4\eta^{2}+5\eta^{4}+5\eta^{6}+4\eta^{8}+3\eta^{10})\, q^{5}+\cdots, \\
\I^{U(5)}_{\rm F}-\I^{U(\infty)}_{\rm F} &= -\frac{1}{\eta^{5}}(1+\eta^{2}+\eta^{4}+\eta^{6}+\eta^{8}+\eta^{10})\, q^{5}\lp
-\frac{1}{\eta^{6}}(3+4\eta^{2}+5\eta^{4}+5\eta^{6}+5\eta^{8}+4\eta^{10}+3\eta^{12})\, q^{6}+\cdots, \\
\I^{U(6)}_{\rm F}-\I^{U(\infty)}_{\rm F} &= -\frac{1}{\eta^{6}}(1+\eta^{2}+\eta^{4}+\eta^{6}+\eta^{8}+\eta^{10}+\eta^{12})\, q^{6}\lp
-\frac{1}{\eta^{7}}(3+4\eta^{2}+5\eta^{4}+5\eta^{6}+5\eta^{8}+5\eta^{10}+4\eta^{12}+3\eta^{14})\, q^{7}+\cdots\, .
\ea
The pattern is 
\ba
\la{B.12}
\I^{U(N)}_{\rm F}-\I^{U(\infty)}_{\rm F} &= \frac{\eta}{1-\eta^{2}}(\eta^{N+1}-\eta^{-N-1})\, q^{N}\lp
+\frac{\eta}{1-\eta^{2}}(3\eta^{N+2}+\eta^{N}+\eta^{N-2}-\eta^{2-N}-\eta^{-N}-3\eta^{-2-N})\, q^{N+1}+\cdots+\mc O(q^{2N}).
\ea

\bibliography{BT-Biblio}
\bibliographystyle{JHEP-v2.9}
\end{document}